\documentstyle{article}
\begin{document}
\title {
Density of States and NMR Relaxation Rate in 
Anisotropic Superconductivity with Intersecting
Line Nodes
}
\author {Yasumasa {\sc Hasegawa}  
 }

\maketitle

\begin{center}{
Faculty of Science, 
Himeji Institute of Technology, \\
Kamigori, Akou-gun, Hyogo 678-12.}
\end{center}

\begin{center}
{ July 12, 1996}
\end{center}


\begin{abstract}
{
We show that the density of states in an anisotropic superconductor
with intersecting line nodes in the gap function is
proportional to $E \log (\alpha \Delta_0 /E)$
 for $|E| \ll  \Delta_0$, where $\Delta_0$ is the maximum value of the gap
function and $\alpha$ is constant, while it is
proportional to $E$ if the line nodes do not intersect.
As a result, 
a logarithmic correction appears in the temperature dependence of
the NMR relaxation rate and the specific heat,
which can be observed experimentally.
By comparing with those for the heavy fermion superconductors,
we can obtain information about the symmetry of the gap function.
}
\end{abstract}



Recently, many effects related to
the anisotropic superconductivity have been observed experimentally in heavy
fermion superconductors.~\cite
{MacLaughlin1984,Kitaoka1985,Kohara1987,Matsuda1996,Kohori1996,Kyogaku1993,Kohori1995} 
The density of states near the Fermi surface is determined experimentally by observing
the temperature dependence of
the specific heat ($C$) and
 the relaxation rate ($T_1^{-1}$) of NMR
and NQR.
It is well known that
if the density of states, $N(E)$, is proportional to
$E^n$,   $1/T_1 \propto T^{2n+1}$ and $C \propto T^{n+1}$.
\cite{Volovik1985,Sigrist1991}

Although the symmetry of the gap function 
  in heavy-fermion superconductors 
has not been determined,  
$T_1^{-1} \propto T^3$ has been observed in NMR and NQR experiments for 
$T < T_c$
in heavy fermion superconductors
(UBe$_{13}$, 
CeCu$_2$Si$_2$, 
UPt$_3$, 
URu$_2$Si$_2$,
and
UPd$_2$Al$_3$).~\cite
{MacLaughlin1984,Kitaoka1985,Kohara1987,Matsuda1996,Kohori1996,Kyogaku1993,Kohori1995} 
This suggests the
existence of line nodes in the gap function $\Delta ({\bf k})$.
\cite{Sigrist1991}
As the simplest model for anisotropic superconductivity with line
nodes in the gap function at the Fermi surface, the $p$-wave
 polar state has been
extensively studied. The gap function for the polar state is given by
\begin{equation}
\Delta_{{\bf k},{\rm polar}} = \Delta_0 \cos \theta,
\end{equation}
where $\cos \theta = k_z / |{\bf k}|$. 
The  effects of impurities  has also been studied
theoretically for the $p$-wave
polar and axial states.~\cite{UedaRice1985,SchmitRink1986,Hirshfeld1986}
However, 
the $p$-wave state is not consistent with
the decrease in the Knight shift below $T_{\rm c}$ 
observed in 
UPd$_2$Al$_3$,~\cite{Kyogaku1993,Kohori1995}
 which indicates that
the pairing has  even parity in that system.
For  even parity pairing we have to consider the $d$-wave or higher wave
pairing state.
If we consider  higher wave paring states with line nodes in the
gap function on the  Fermi surface, 
 the line nodes intersect in some cases. 
 For example,
the line nodes intersect at $(0, 0, \pm k_{\rm F})$ on the Fermi surface 
for the state  with the $d$-wave 
gap function, $\Delta_{\bf k} \propto k_x^2
- k_y^2$ or $\Delta_{\bf k} \propto k_x k_y$.
Although  intersection of the line nodes on the Fermi surface may not be
favored by the condensation energy, it is not forbidden by the
symmetry.

 $T^3$ dependence of $T_1^{-1}$ has also been observed in high $T_{\rm c}$
superconductors~\cite{Imai1988,Kitaoka1988}
 and organic superconductors.~\cite{Takigawa1987,Mayaffre1995}
Anisotropic
superconductivity with line nodes is also thought to be realized 
in these materials.
Since
these superconductors have low-dimensional Fermi surfaces, 
the line nodes in the gap function are not likely to intersect on the Fermi
surface.

It is often stated that
$N(E) \propto E^2$ in the case of point nodes and
$N(E) \propto E$ in the case of line nodes.~\cite{Volovik1985,Sigrist1991}
However, it has not been noticed, as far as the author knows, that
$N(E) \propto E$  only when line nodes do not intersect.
In this work we show 
that
when the gap function is zero on the intersecting lines,
$N(E) \propto E \log (\alpha \Delta_0 /E)$, where
$\alpha$ is constant and $\Delta_0$ is the maximum value of the gap function.
 As a result, the
temperature dependence of $T_1^{-1}$  
 differs significantly from that
for the case with non-intersecting line nodes. 

We assume a spherical Fermi surface. 
The density of states is given by
\begin{eqnarray}
N(E) 
 &=& N_0 \int \frac{{\rm d} \Omega}{4 \pi} 
  {\rm Im} \frac {E}{\sqrt{\Delta_{\bf k}^2-E^2}} ,
\end{eqnarray}
where the density of states for the normal state, $N_0$, is constant
 and the gap function $\Delta_{\bf k}$ depends
only on the direction of ${\bf k}$.

The density of states for the polar state is given by
\begin{equation}
\frac{N_{\rm polar}(E)}{N_0} = 
\left\{
 \begin{array}{ll}
 {\displaystyle \frac{\pi}{2} \frac{E}{\Delta_0} } &
 \mbox{if $E < \Delta_0$} \\
 & \\
 {\displaystyle \frac{E}{\Delta_0} \arcsin \frac{\Delta_0}{E} } &
 \mbox{if $E \geq \Delta_0$}  \\
 \end{array}
\right.         .
\end{equation}

We calculate the density of states for the state
with the gap function given by 
\begin{eqnarray}
\Delta_{{\bf k}, x^2-y^2} &=& \Delta_0 \frac{k_x^2 - k_y^2}{|{\bf k}|^2}
\nonumber \\
&=& \Delta_0 \sin^2 \theta \cos 2 \phi,
\end{eqnarray}
or
\begin{eqnarray}
\Delta_{{\bf k}, xy} &=& \Delta_0 \frac{2 k_x k_y}{|{\bf
k}|^2}
\nonumber \\
&=& \Delta_0 \sin^2 \theta \sin 2 \phi.
\end{eqnarray}
These two states are degenerate since we assume a spherical Fermi surface.
As shown in Fig.1 the gap function for the $x^2 - y^2$ state
 is zero on the lines $k_x = k_y$ and
$k_x=-k_y$ 
on the Fermi surface
and these lines intersect at points $(0,0, \pm k_{\rm F})$.
The density of states is obtained as
\begin{eqnarray}
\frac{N_{x^2-y^2}(E)}{N_0} 
 & & =
\left\{   
\begin{array}{ll}
 {\displaystyle
  \frac{x}{\pi } 
    \left[ \int_0^1 \frac{K(u)}{\sqrt{1-x u}} {\rm d} u  \right. } & \\
 \hspace*{0.5cm}
 {\displaystyle
 \left.
    + \int_0^{1-x} \frac{K(u+x)}{\sqrt{u (u+x)}} {\rm d} u   \right] }
  & \mbox{if {\it x} $<$ 1} \\
  &  \\
 {\displaystyle 
  \frac{x}{\pi} \int_0^{\frac{1}{x}} \frac{K(u)}{\sqrt{1 - x u}} {\rm d} u  }
  & \mbox{if {\it x} $\geq$ 1} \\
  \end{array}
\right.
\end{eqnarray}
where $x = E / \Delta_0$ and
 $K(k)$ is the complete elliptic integral of the first kind given by
\begin{equation}
K(k) = \int_0^{\frac{\pi}{2}} \frac{1}{\sqrt{1-k^2 \sin^2 \theta}} {\rm d}
\theta .
\end{equation}
For $E \ll \Delta_0$, we obtain
\begin{equation}
\frac{N_{x^2-y^2}(E)}{N_0} \approx \frac{E}{2 \Delta_0} 
 \log \frac {\alpha \Delta_0}{E},
\label{dos}
\end{equation}
where  $\alpha$ is obtained numerically as
$\alpha = 16.0$.
We plot $N(E)$ for the polar  and $x^2-y^2$ states
 as a function of $E$ in Fig.~2. 
The densities of states for various states were calculated numerically
by
Hasselbach {\it et al}.,~\cite{Hasselbach1993} 
but they did not notice the logarithmic dependence. 

The $E \log (\alpha \Delta_0 / E)$ dependence of the density of 
states is a general
property of anisotropic superconductors with intersecting line
nodes in the gap function on the Fermi surface. Near the points of intersection of
line nodes, we can write the gap function as
\begin{equation}
\Delta_{\bf k} \sim \Delta_0 \xi \eta,
\end{equation}
by choosing  suitable variables $\xi$ and $\eta$ on the
Fermi surface. For $x=E/\Delta_0 \ll 1$, the density of states
due to these line nodes is obtained as
\begin{eqnarray}
N(E) &\sim& N_0 \int_{-1}^1 {\rm d}\xi \int_{-1}^1 {\rm d}\eta {\rm Im} 
 \frac{E}{\sqrt{(\Delta_0\xi\eta)^2-E^2}} \nonumber \\
 &=&4 N_0 x \left[ \int_0^x {\rm d}\xi \int_0^1 {\rm d} \eta
\frac{1}{\sqrt{x^2-(\xi\eta)^2}} \right. \nonumber \\
& & \left. + \int_x^1 {\rm d}\xi \int_0^{\frac{x}{\xi}} {\rm d}\eta 
\frac{1}{\sqrt{x^2-(\xi\eta)^2}} \right] \nonumber \\
&=& 2 \pi N_0 \frac{E}{\Delta_0} \log \frac{2\Delta_0}{E},
\end{eqnarray}
as obtained in eq.(\ref{dos}) for the $x^2-y^2$ state.

In order to show that  $N(E) \propto E \log (\alpha\Delta_0/E)$
can be observed  experimentally, we calculate 
the NMR relaxation rate $T_1^{-1}$, which is given by
\begin{eqnarray}
{\frac{(T_1)_{\rm normal}}{(T_1)_{\rm super} }  } 
&=&
 2  \int_0^{\infty}
 \left( - \frac{\partial f (E)}{\partial E} \right)
\nonumber \\
& & \times
 \left\{ 
\left( \frac{N(E)}{N_0} \right)^2 + \left( \frac{M(E)}{N_0}
 \right)^2 \right\} {\rm d}E,
\label{eqT1}
\end{eqnarray}
where
\begin{equation}
M(E) = N_0 \int \frac{{\rm d} \Omega}{4 \pi} {\rm Im} \frac{\Delta_{\bf k}}{\sqrt{
\Delta_{\bf k}^2 - E^2}}  .
\end{equation}
For polar and $x^2-y^2$ states, for which
 the $s$-wave component is not included  in the gap
function, we obtain $M(E) = 0$.
We calculate eq.(\ref{eqT1}) numerically by assuming the temperature
dependence of the maximum gap to be
\begin{equation}
\Delta_0 (T) = 1.76 k_{\rm B} T_{\rm c} 
   \tanh \left( 1.74 \sqrt{ \frac{T_{\rm c}}{T} - 1} \right) ,
\label{approximate}
\end{equation}
which is the approximate temperature dependence of the energy gap
 in  $s$-wave
superconductors in the weak coupling limit.  
Since we are interested in the effect of the logarithmic correction in the
density of states due to the intersection of the line nodes in the gap
function, we use eq.(\ref{approximate})  for the polar
 and $x^2 - y^2$ states.

In Fig.~3 we plot $T_1^{-1}$ normalized by the value for the normal state
as a function of temperature. 
In order to see the temperature dependence
clearly we plot $1 / (T_1 T^3)$ as a function of temperature in Fig.~4,
which shows that while
a $T^3$ dependence of $ T_1^{-1}$ is obtained for $T \leq
 T_{\rm c} /2$ for the polar state, 
$1/(T_1T^3)$ diverges as the temperature decreases
for the $x^2-y^2$ state.
At low temperatures we obtain
$T_1^{-1}\propto T^3 (\log T )^2$ for the $x^2-y^2$
state.
 
The experimental results for the heavy fermion superconductors 
UBe$_{13}$,~\cite{MacLaughlin1984} 
CeCu$_2$Si$_2$,~\cite{Kitaoka1985}
URu$_2$Si$_2$~\cite{Kohara1987,Matsuda1996,Kohori1996} and
UPd$_2$Al$_3$~\cite{Kyogaku1993,Kohori1995} 
are
well fitted by $T_1^{-1} \propto T^3$ and no logarithmic correction
is observed. 
It can be concluded that the line nodes
in the gap function do not intersect on the Fermi surface. 
This provides a   restriction for the symmetry of the gap function
and the Fermi surface. If the gap function has the symmetry with  
$k_x^2-k_y^2$ or $k_x k_y$, the Fermi surface should open in the 
$(0,0,\pm 1)$ direction, as in the quasi-one or quasi-two
dimensional case. 

The temperature dependence of the specific heat for 
various  anisotropic
superconductors has been calculated 
by Hasselbach {\it et al}.~\cite{Hasselbach1993} 
They obtained  the deviation from  linear
dependence in the density of states and $C(T)/T$ for the states with
intersecting line nodes numerically, although
  they did not observe the logarithmic corrections.
From the temperature dependence of the specific heat, 
it can be concluded that intersection of the line nodes
does not occur on the Fermi surface. 
  
In conclusion,  we have shown that intersection 
of the line nodes results in an $E
\log (\alpha \Delta_0 /E)$ dependence in the density of states.
In that case deviation from 
power law behavior can be observed in the temperature dependence of
the relaxation rate $T_1^{-1}$ and
the specific heat. 
In  experiments on heavy fermion superconductors, no deviation
from  power law dependence has been observed. 
This implies a the strong
restriction on the symmetry of the gap function, i.e. the gap function
should not have  saddle points on the Fermi surface.

The author would like to thank Professor Y. Kohori for   discussion
on the experimental results and Professor T. Kohara for valuable
comments.


\newpage
\begin{figure}
\caption{
The gap function
 for the $x^2 - y^2$ state 
}
\label{Fig.1}
\end{figure}

\begin{figure}
\caption{
The densities of states for the $x^2 - y^2$ state (solid line) 
and the polar state
(broken line).
}
\label{Fig.2}
\end{figure}
\begin{figure}
\caption{
The NMR relaxation rates 
${\left( T_1 \right)_{\rm normal}} / {\left( T_1 \right)_{\rm super}}$
for the $x^2 - y^2$ state (solid line) and the polar state (broken line)
 as a function of temperature.
}
\label{Fig.3}
\end{figure}
\begin{figure}
\caption{
The NMR relaxation rates, 
${\left( T_1 T^3 \right)_{T=T_{\rm c}}}/{\left( T_1 T^3 \right)_{\rm super}}$,
for the $x^2 - y^2$ state (solid line) and the polar state (broken line)
 as a function of temperature.
}
\label{Fig.4}
\end{figure}

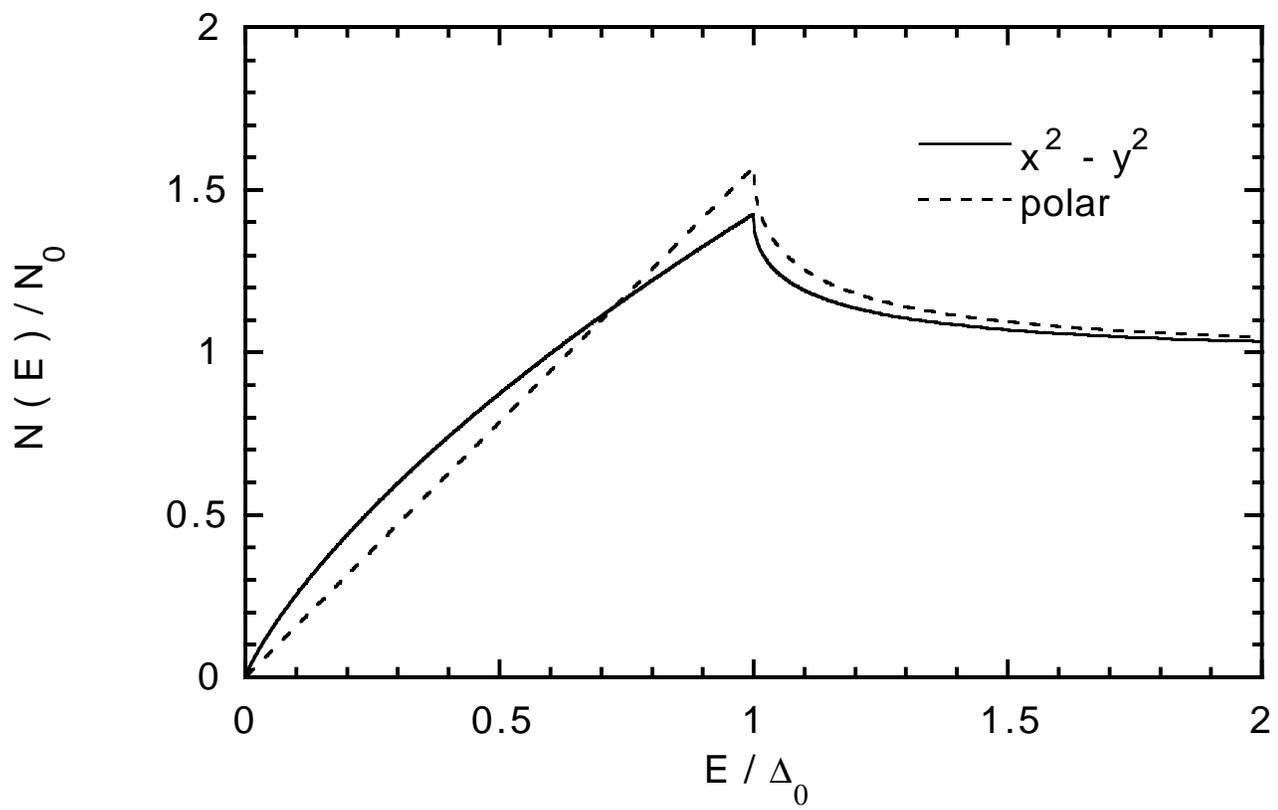
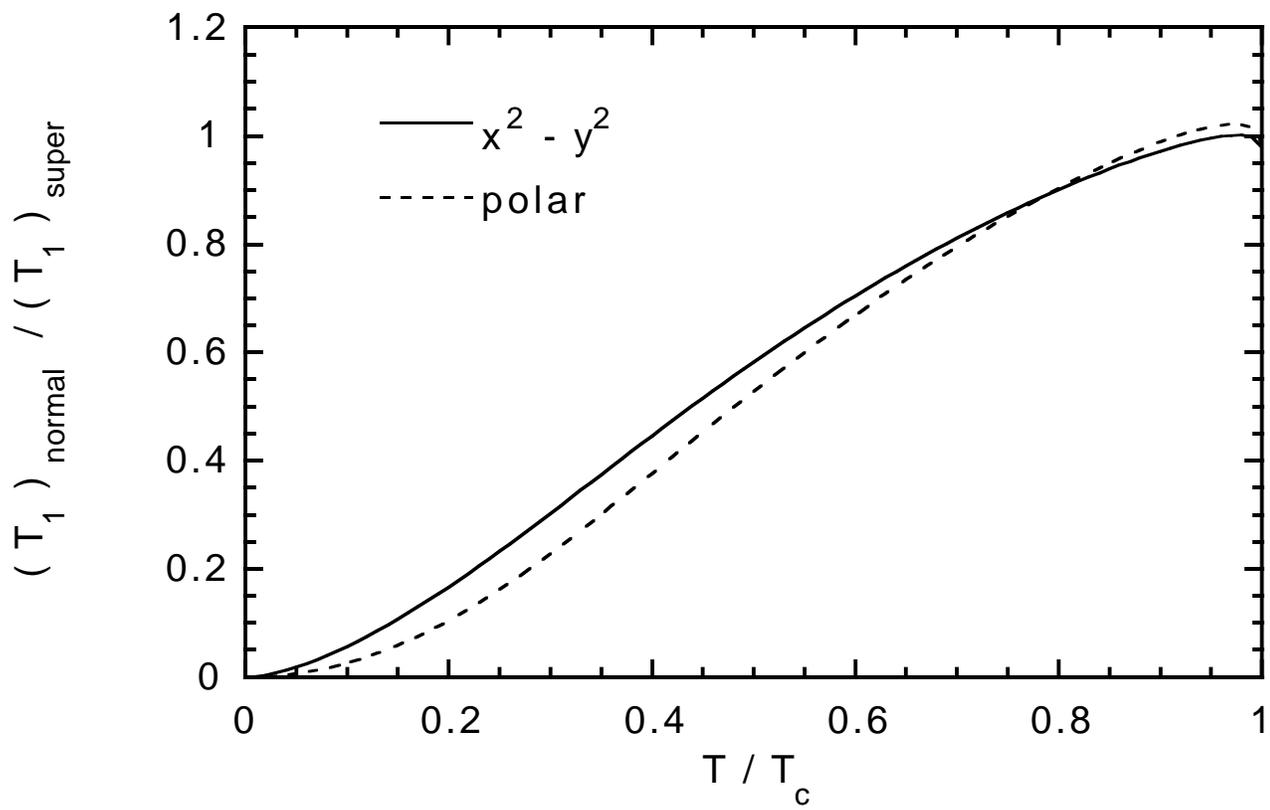
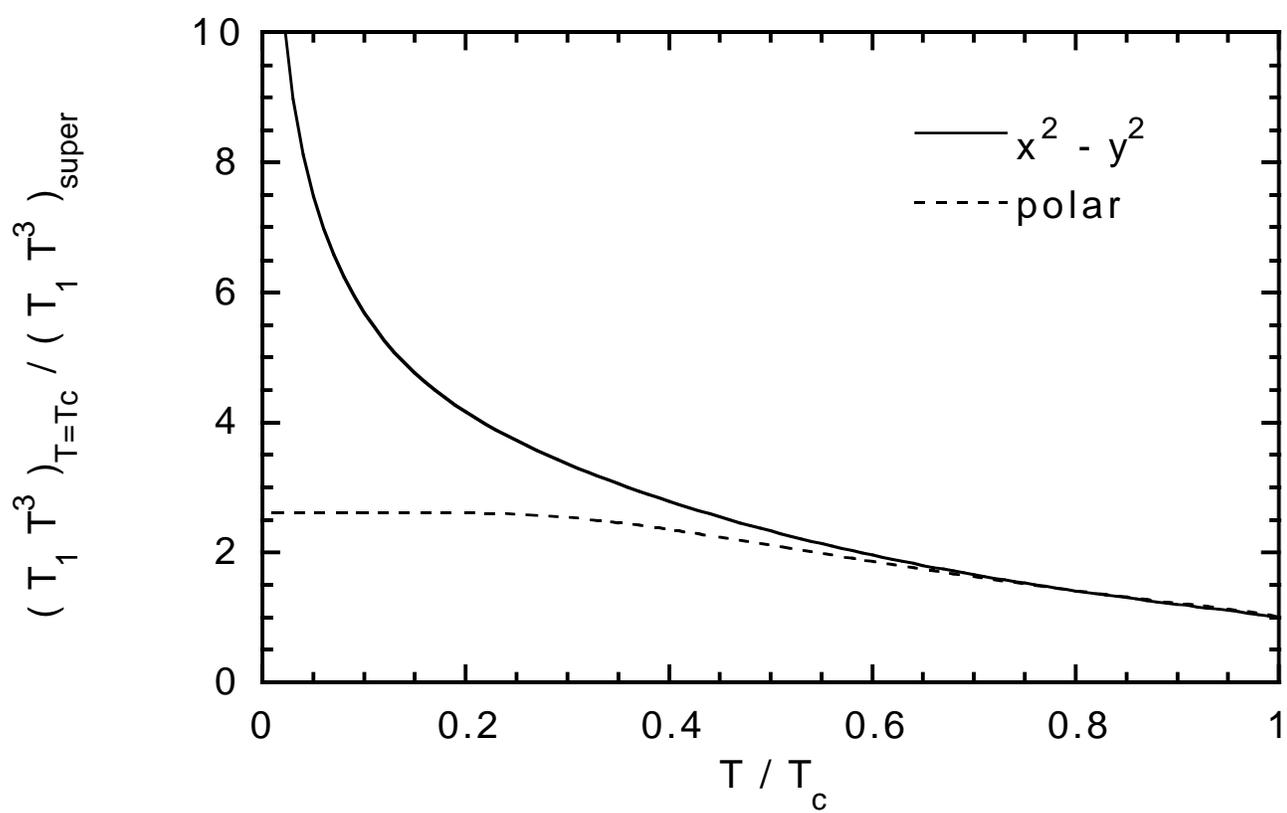
\end{document}